\newif\ifarxiv
\arxivtrue

\documentclass[cmp,final]{svjour}
\usepackage{cite}
\usepackage{amsmath}
\usepackage{amsfonts}
\usepackage{amssymb}
\usepackage{epsf}
\usepackage{graphicx}
\usepackage{graphics}
\usepackage{verbatim}
\usepackage{epsfig}

\IfFileExists{myowntimes.sty}
	{\usepackage{myowntimes}}
	{\usepackage{times}\usepackage{mathrsfs}}
	
\DeclareFontFamily{OT1}{eusm}{} \DeclareFontShape{OT1}{eusm}{m}{n} {<5> <6> <7> <8> <9> <10> <11> <12> <14.4> eusm10}{}
\DeclareMathAlphabet{\eusm}{OT1}{eusm}{m}{n}

\DeclareFontFamily{OT1}{fraktura}{}
\DeclareFontShape{OT1}{fraktura}{m}{n} {<5> <6> <7> <8> <9> <10> <11> <12> <13> <14.4> [1.1] eufm10}{}
\DeclareMathAlphabet{\fraktura}{OT1}{fraktura}{m}{n}

\ifarxiv
\fi

\newenvironment{proofsect}[1]{\vskip0.1cm\noindent{\rmfamily\itshape #1.}}{\qed\vspace{0.15cm}}

\spnewtheorem{mylemma}[theorem]{Lemma}{\bf}{\it}
\spnewtheorem{myproposition}[theorem]{Proposition}{\bf}{\it}
\spnewtheorem{mycorollary}[theorem]{Corollary}{\bf}{\it}
\spnewtheorem{mydefinition}[theorem]{Definition}{\bf}{\it}
\spnewtheorem{myremark}[theorem]{Remark}{\it}{\rm}
\spnewtheorem{myremarks}[theorem]{Remarks}{\it}{\rm}
\spnewtheorem{myproblem}{Problem}{\bf}{\it}
\numberwithin{equation}{section}
\numberwithin{theorem}{section}

\newcommand{\dist}{\operatorname{dist}}

\newcommand{\textd}{\text{\rm d}\mkern0.5mu}

\newcommand{\texte}{\text{\rm e}}

\renewcommand{\AA}{\mathcal A}

\newcommand{\EE}{\mathcal E}

\newcommand{\QQ}{\mathcal Q}

\newcommand{\R}{\mathbb R}

\newcommand{\T}{\mathbb T}

\newcommand{\Z}{\mathbb Z}

\newcommand{\twoeqref}[2]{(\ref{#1}-\ref{#2})}
\newcommand{\cc}{{\text{\rm c}}}

\def\myffrac#1#2 in #3{\raise 2.6pt\hbox{$#3 #1$}\mkern-1.5mu\raise 0.8pt\hbox{$#3/$}\mkern-1.1mu\lower 1.5pt\hbox{$#3 #2$}}
\newcommand{\ffrac}[2]{\mathchoice%
	{\myffrac{#1}{#2} in \scriptstyle}
	{\myffrac{#1}{#2} in \scriptstyle}
	{\myffrac{#1}{#2} in \scriptscriptstyle}
	{\myffrac{#1}{#2} in \scriptscriptstyle}
}

\begin{document}

\title{On the absence of ferromagnetism in typical 2D ferromagnets}
\titlerunning{Absence of ferromagnetism in 2D ferromagnets}
\author{Marek Biskup\inst{1}\and Lincoln Chayes\inst{1}\and Steven~A. Kivelson\inst{2}}
\authorrunning{M.~Biskup, L.~Chayes, S.A.~Kivelson}
\institute{Department of Mathematics, UCLA, Los Angeles, CA 90095-1555, U.S.A.\and Department of Physics, Stanford University, Stanford, CA 94305-4045, U.S.A.}
\date{}
\maketitle

\renewcommand{\thefootnote}{}
\footnotetext{\vglue-0.41cm\footnotesize\copyright\,2006 by M.~Biskup, L.~Chayes and S.A.~Kivelson. Reproduction, by any means, of the entire article for non-commercial purposes is permitted without charge.}
\renewcommand{\thefootnote}{\arabic{footnote}}

\begin{abstract}
We consider the Ising systems in~$d$ dimensions with nearest-neighbor ferromagnetic interactions and long-range repulsive (antiferromagnetic) interactions that decay with power~$s$ of the distance. The physical context of such models is discussed; primarily this is $d=2$ and~$s=3$ where, at long distances, genuine magnetic interactions between genuine magnetic dipoles are of this form. We prove that when the power of decay lies above $d$ and does not exceed $d+1$, then for all temperatures the spontaneous magnetization is zero. In contrast, we also show that for powers exceeding~$d+1$ (with $d\ge2$) magnetic order can occur.
\end{abstract}

\section{Introduction}
\label{sec1}\noindent
While most of our knowledge of statistical mechanics is derived from studies of model problems with short-range forces, in nature interactions more often fall off only in proportion to an inverse power of the distance, $U(r) \sim 1/r^s$.  This includes systems interacting via Coulomb forces (\hbox{$s=1$}), dipolar interactions ($s=3$) as well as interactions caused by collective effects such as strain induced interactions in solids or the effective entropic interactions (analogous to Casimir forces) in lipid films.  
When the interactions are sufficiently long-range, i.e.,~when  $s \leq d$ where~$d$ is the spatial dimension,  the very definition of the thermodynamic limit is different than for short-ranged models.  
However, even when \hbox{$s > d$} there can be qualitatively new, or at least unexpected, phenomena, cf, e.g.,~\cite{Thouless,AY,ACCN,AN}.

In the present paper we study a class of systems with long-range forces; namely, the Ising models on $\Z^d$, $d\ge1$, which are defined by the (formal) Hamiltonians
\begin{equation}
\label{Hamiltonian}
H = -\sum_{\langle i,j \rangle}J\sigma_{i}\sigma_{j}
+ \frac12\sum_{i,j}K_{i,j}\sigma_{i}\sigma_{j}.
\end{equation}
Here $\sigma_{i} \in \{+1,-1\}$, $i$ and $j$ index sites in $\Z^d$ and $\langle i,j \rangle$ denotes a nearest neighbor pair. The above notation expresses the relevant signs of all the couplings:  $J > 0$ is the short-range \emph{ferromagnetic} interaction while $K_{i,j} \geq 0$ represents the \emph{antiferromagnetic} long range interaction which we assume decays with power~$s$ of the distance between~$i$ and~$j$. We investigate the question of presence, and absence, of spontaneous magnetization in such models.  

The motivation for this work was provided by a paper of Spivak and one of us~\cite{SK} where it was conjectured that, in the presence (or absence) of an external field, discontinuous transitions permitting coexisting states of different magnetization are forbidden for antiferromagnetic power law interactions with range $d < s \le d+1$. A heuristic proof by contradiction was presented based on the explicit construction of a ``micro-emulsion'' phase which has a lower free energy than the state of macroscopic two-phase coexistence. Simply put, the anticipated surface tension between the two pure phases would be negative---and divergent.  The proof is heuristic in the sense that it makes the physically plausible assumption that correlations in the putative coexisting phases have reasonable decay and that there is a well defined interface.

As it turns out, versions of the above conjecture are actually more than 20 years old. For example, on the physics side, modulated phases in 2D dipolar ferromagnets were analyzed in \cite{doniach,pokrovsky1,pokrovsky2}. On the mathematics side, in~\cite{vanEnter}, models with extreme anisotropic repulsive interactions which have \emph{very} slow decay, but only among a sparse set of spins, were considered and absence of spontaneous magnetism was proved. The isotropic case, $U(r)\sim1/r^s$, was also mentioned in~\cite{vanEnter} and the significance of the interval $d<s<d+1$ for the absence of magnetization was highlighted (with no mention of~$s=d+1$). Related problems were described in~\cite{JKS} for systems with longer range, e.g., Coulomb, interactions and in \cite{vanEnterPRB,vanEnter} for the current setup with $O(n)$-spins. Furthermore, general theorems demonstrating instability of phase coexistence under the addition of generic long-range interactions have been proved in~\cite{Daniels-vE,Israel,Sokal}. In the present paper we provide a full proof of the absence of ferromagnetism in the model \eqref{Hamiltonian} with~$d<s\le d+1$ thereby vindicating completely the arguments of~\cite{SK}---at least for $h=0$.

The mathematical result presented in this note has the following consequence for 2D physics:  Two-dimensional magnetic systems often have strong ``crystal field'' effects which orient the electron spins (largely or entirely) in the $z$ direction, perpendicular to the plane in which they reside.  This gives the problem of magnetic ordering an Ising character.
Interactions between nearby spins---quantum mechanical and somewhat complicated---are, often enough, of the ferromagnetic type and considerably stronger than the direct magnetic dipolar interactions (which are a relativistic effect).  Thus, it seems reasonable to study Ising ferromagnets in 2D contexts and conclude that there is a definitive possibility for ferromagnetism.  However, while possibly weak, there
is \textit{always} the long-range~$1/r^3$ repulsive interaction. The conclusion of this note is that, no matter how small its relative strength may be, this interaction will preclude the possibility of ferromagnetism among the $z$-components.

We remark that the absence of magnetization certainly does not disallow other types of ordering. Indeed, a large body of physics literature
\cite{doniach,pokrovsky1,pokrovsky2,SK,schlovsky1,schlovsky2,vanderbilt,low,romans,chayes,tarjus,bak}, points in the direction of modulated (striped and/or bubble) states in this and related systems. (For an extremely insightful review of the phases produced by models of this sort and many experimentally clear realizations of the corresponding physics, see~\cite{seul}.) From the perspective of mathematics, recent rigorous estimates on ground-state energies~\cite{GLL}, which are asymptotic in $d\ge2$ and exact in $d=1$, also indicate striped order in the ground state. In fact, for certain special cases of the 1D ground-state problem, this has been established completely.

The organization of the rest of this paper is as follows: In the next section we define all necessary background and state the main results. In Sect.~\ref{sec3} we derive some estimates on the strength of the long-range interaction between a box and its complement. These are assembled into the proof of the main result in Sect.~\ref{sec4}. Sect.~\ref{sec5} contains some open problems and further discussion.

\section{Statement of main results}
\label{sec2}\noindent
As mentioned, for the problem of central interest we have $K_{i,j}\sim|i-j|^{-3}$ in~$d=2$, where $\vert i-j\vert$ is the Euclidean distance, but we may as well treat all powers for which the interaction is absolutely summable.  To be definitive we will simply take, for $s>d$,
\begin{equation}
\label{K term}
K_{i,j} = \frac{1}{\vert i-j\vert^{s}}
\end{equation}
with the proviso~$K_{i,i}=0$.  We remark that more generality than  \eqref{K term}
is manifestly possible as is also the case with the ferromagnetic portion of the interaction in  \eqref{Hamiltonian}.  However, these generalities would tend to obscure the mechanics of the proofs and so we omit them.

In order to define the corresponding Gibbs measures, let~$\Lambda\subset\Z^d$ be a finite set and, given a configuration $\sigma\in\{+1,-1\}^{\Z^d}$, let~$H_\Lambda=H_\Lambda(\sigma_\Lambda,\sigma_{\Lambda^\cc})$ denote the Hamiltonian in~$\Lambda$ which is obtained from \eqref{Hamiltonian} by pitching out all terms with \emph{both}~$i$ and~$j$ outside~$\Lambda$. Since~$s>d$, the corresponding object is bounded uniformly in~$\sigma$. Then the DLR formalism tells us that a probability measure on~$\{+1,-1\}^{\Z^d}$---equipped with the product $\sigma$-algebra---is a Gibbs measure if the regular conditional distribution of~$\sigma_\Lambda=(\sigma_i)_{i\in\Lambda}$ given a configuration $\sigma_{\Lambda^\cc}=(\sigma_i)_{i\in\Lambda^\cc}$ in the complement $\Lambda^\cc=\Z^d\setminus\Lambda$ is of the form
\begin{equation}
Z_\Lambda(\sigma_{\Lambda^\cc})^{-1}\texte^{-\beta H_\Lambda(\sigma_\Lambda,\sigma_{\Lambda^\cc})},
\end{equation}
where
\begin{equation}
Z_\Lambda(\sigma_{\Lambda^\cc})=\sum_{\sigma_\Lambda}\texte^{-\beta H_\Lambda(\sigma_\Lambda,\sigma_{\Lambda^\cc})}
\end{equation}
is the partition function. We will use the notation $\langle-\rangle$ to denote expectations with respect to Gibbs measures (which may often stay implicit).

We wish to establish that all Gibbs measures corresponding to the above Hamiltonian have zero average magnetization once~$s\in(d,d+1]$. We will employ some thermodynamic arguments based, ultimately, on the notion of the free energy. To define this quantity, let~$Z_{\Lambda,h}(\sigma_{\Lambda^\cc})$ 
denote the partition function in $\Lambda$ with the Hamiltonian
\begin{equation}
H_\Lambda(\sigma_\Lambda,\sigma_{\Lambda^\cc})-h\sum_{i\in\Lambda}\sigma_i,
\end{equation}
i.e., for the model in homogeneous external field~$h$. Let
\begin{equation}
\label{Lambda-L}
\Lambda_L=[-L,L]^d\cap\Z^d.
\end{equation}
Then there exists~$\epsilon_L=o(|\Lambda_L|)$---with little-$o$ uniform in~$h$---such that for all $\sigma,\tilde\sigma\in\{-1,1\}^{\Z^d}$,
\begin{equation}
\label{surface-term}
\biggl|\log\frac{Z_{\Lambda_L,h}(\sigma_{\Lambda_L^\cc})}{Z_{\Lambda_L,h}(\tilde\sigma_{\Lambda_L^\cc})}\biggr|
\le\epsilon_L.
\end{equation}
In particular, the limit
\begin{equation}
\label{pressure}
f(\beta,h)=-\frac1\beta\,\lim_{L\to\infty}\frac1{|\Lambda_L|}\log Z_{\Lambda_L,h}(\sigma_{\Lambda_L^\cc})
\end{equation}
exists and is independent of the boundary condition. Furthermore, the function~$h\mapsto f(\beta,h)$ is concave for all~$h$. 

The independence of the free energy on the boundary condition is standard and follows from the uniform bound on energy per site; cf \cite[Theorem~II.3.1]{Simon}. In Sect.~\ref{sec3} we will show that, perhaps not surprisingly,~$\epsilon_L$ is order~$L^{\max\{2d-s,d-1\}}$ with a logarithmic correction at~$s=d+1$.

The concavity of the free energy now permits us to define the \emph{spontaneous magnetization} $m_\star=m_\star(\beta)$ via the right-derivative of $h\mapsto f(\beta,h)$ at~$h=0$:
\begin{equation}
\label{spont-mag}
m_\star=-\frac{\partial f}{\partial h^+\!}\,\Bigl|_{h=0}.
\end{equation}
It is clear that, by the plus-minus symmetry built into the model, the corresponding left derivative equals~$-m_\star$.

\smallskip
The statement of our main result is as follows:

\begin{theorem}
\label{main theorem}
Consider the interaction described by the Hamiltonian in \twoeqref{Hamiltonian}{K term}. Then for all $s\in(d,d+1]$ and all $\beta\in(0,\infty)$, the spontaneous magnetization,~$m_\star$, is zero.
\end{theorem}

The regime~$d<s<d+1$ of exponents for the vanishing of the spontaneous magnetization was surmised already in~\cite{vanEnter}; the present work covers this and, in addition, the somewhat subtle borderline case $s=d+1$. The above is about as strong a statement as possible concerning the absence of magnetic order from a thermodynamic perspective; the implications for statistical mechanics are similar in their finality.  Indeed, the following standard conclusions are implied for the properties of equilibrium states:

\begin{mycorollary}
\label{big-D}
Let~$s\in(d,d+1]$ and let $\mu$ be any infinite-volume Gibbs measure for the Hamiltonian in \twoeqref{Hamiltonian}{K term} at inverse temperature~$\beta\in(0,\infty)$. Let $\Lambda_L$ be as in \eqref{Lambda-L}. For each $\epsilon>0$ there exists~$\delta>0$ such that for all $L$ sufficiently large and $\mu$-almost every boundary condition~$\sigma_{\Lambda_L^\cc}$,
\begin{equation}
\mu\biggl(\,\Bigl|
\sum_{i \in \Lambda_L}\sigma_i\Bigr|>\epsilon|\Lambda_L|\bigg|\sigma_{\Lambda_L^\cc}\biggr)\le\texte^{-\delta|\Lambda_L|}.
\end{equation}
In particular, $\mu$-almost all configurations~$\sigma$ have zero block-average magnetization,
\begin{equation}
\lim_{L\to\infty}
\frac1{|\Lambda_L|}
\sum_{i \in \Lambda_L}\sigma_i = 0.
\end{equation}
Finally, in any translation-invariant (infinite volume) Gibbs state, the expectation of the spin at the origin is zero.
\end{mycorollary}

The last statement should not be interpreted as a claim that the state is disordered. In fact, as already mentioned, one expects the occurrence of ``striped states'' at sufficiently low temperatures; see our discussion in Sect.~\ref{sec1} and also Sect.~\ref{sec5}. Note that no restrictions are put the nearest-neighbor coupling~$J$; the theorem works for all~$J\in\R$. 

To complement our ``no-go'' Theorem~\ref{main theorem}, we note that for exponents~$s>d+1$, spontaneous magnetization \emph{will} occur under the ``usual'' conditions:

\begin{theorem}
\label{thm-peierls}
Let~$d\ge2$, pick $s>d+1$ and consider the interaction as described in \twoeqref{Hamiltonian}{K term}. Then there exist$J_0=J_0(s,d)\in(0,\infty)$ and~$C_0=C_0(d)\in(0,\infty)$ such that for all~$\beta(J-J_0)\ge C_0$,
\begin{equation}
m_\star>0.
\end{equation}
In particular, under such conditions, there exist two distinct, translation-invariant extremal Gibbs states~$\langle-\rangle^+$ and~$\langle-\rangle^-$ such that
\begin{equation}
\langle\sigma_0\rangle^+=-\langle\sigma_0\rangle^->0.
\end{equation}
\end{theorem}

Strictly speaking, this result could be proved by directly plugging in a theorem from~\cite[Section~3]{Ginibre}, which is based on an enhanced Peierls estimate. Instead, we provide an independent way to estimate the contour-flip energy which is technically no more demanding and permits the use of sharp contour-counting arguments~\cite{Lebowitz-Mazel} to derive good estimates on~$J_0$ and the critical value of~$\beta$ at which the transition occurs. As a result, the corresponding constants can be bounded as follows: 
\begin{equation}
J_0\le C\frac{\pi_d}{d+1-s}\quad\text{and}\quad
C_0\le C\frac{\log d}d,
\end{equation}
where~$\pi_d$ is the ``surface'' measure of the unit sphere in~$\R^d$, and $C$ is a constant of order unity.

\section{Estimates on interaction strength}
\label{sec3}\noindent
In this section we will perform some elementary but in places tedious calculations that are needed for the proof of our main results. We begin by an estimate on the energy cost of turning large magnetized blocks to opposite magnetization:

\begin{myproposition}
\label{total}
Let $\Lambda_{L}$ be as above and, for the couplings $K_{i,j}$ described in  \eqref{K term}, consider the discrete sum 
\begin{equation}
T_{L} =
\sum_
{\begin{subarray}
{c} i \in \Lambda_{L} \\ j \in \Lambda^\cc_L
\end{subarray}}
 K_{i,j}.
\end{equation}
Then, as $L$ tends to infinity: 
\settowidth{\leftmargini}{(111)}
\begin{enumerate}
\item[(i)] 
For $d < s < d+1$,
\begin{equation}
T_{L} \sim L^{2d-s}\mathcal Q
\end{equation}
where $\mathcal{Q}\in(0,\infty)$ is the integral
\begin{equation}
\label{Integral Q}
\mathcal{Q}  =  \int _
{\begin{subarray}
{c} x \in S_{1} \\ y \in S_{1}^\cc
\end{subarray}}
 \frac{\textd x\,\textd y}{\vert x - y\vert ^{s}}.
\end{equation}
with $S_{1}=\{x\in\mathbb R^{d}\colon|x|_1\le1\}$.
\item[(ii)]
For $s = d+1$, there exists a constant~$A=A(d)\in(0,\infty)$ such that
\begin{equation}
 \label{LlogL}
T_{L} \sim AL^{d-1}\log L.
\end{equation}
\end{enumerate}
In both (i) and (ii) the symbol $\sim$ is interpreted to mean that the ratio of the two sides tends to unity in the stated limit.
\end{myproposition}

To prove this claim, we will instead consider the quantity $T_{L,a}$ which is defined in the same fashion as $T_L$ except that the ``inside sum'' now ranges over $\Lambda_{L-a}$ instead of 
$\Lambda_{L}$, providing us with a cutoff scale $a$. Of course we must allow $a \to \infty$ and, for $s \in (d,d+1)$, not much more is actually required but, to save work, we shall insist that $a/L^{d+1-s} \to 0$. (Indeed, we remark that while most of the up and coming is not \emph{strictly} necessary for these cases, it will allow for a unified treatment later.) For the the marginal case of $s = d+1$ we need to implement the stronger requirement that $a/\log L \to 0$.

Our claim is that the augmented quantities have the asymptotics that was stated for their unadorned counterparts.  This is sufficient since, keeping in mind the above requirements,
\begin{equation}
T_{L,a}  \leq  T_L  \leq T_{L,a} + 2d\EE_sL^{d-1}a
\end{equation}
where
\begin{equation}
\EE_s=\sum_j K_{0,j} < \infty
\end{equation}
denotes the maximum antiferromagnetic energy associated with a single spin flip.

For the purposes of explicit calculations, it will be convenient to replace $K_{i,j}$ with the quantities~$\tilde{K}_{i,j}$ obtained by ``smearing'' the interaction about the unit cells surrounding the sites~$i$~and~$j$:
\begin{equation}
\tilde{K}_{i,j}  =  
\int _
{\begin{subarray}
{c} \vert x -i\vert _{\infty} \leq \ffrac{1}{2} \\ \vert y - j\vert _{\infty} \leq \ffrac{1}{2}
\end{subarray}}
 \frac{\textd x\,\textd y}{\vert x - y\vert ^{s}}.
\end{equation}
It is noted that since all distances exceed (the large quantity) $a$, the approximation is not severe:
\begin{equation}
\label{KKtilde}
\frac{K_{i,j}}{(1 + \theta a^{-1})^{s}} \leq \tilde{K}_{i,j} \leq 
\frac{K_{i,j}}{(1 - \theta a^{-1})^{s}}
\end{equation}
where $\theta$ is a number of order unity.  Thus, to prove the asymptotics for the $T_{L,a}$, we may insert the 
$\tilde{K}_{i,j}$ and then perform blatant continuum integration.  

\smallskip
As a technical step, for the proof we will need to calculate the total (long-range) interaction between the line segment $(-L,-a)$ on the $x$-axis and the half-space in~$\R^d$ containing all points with positive $x$-coordinate:

\begin{mylemma}
\label{lemma-I1}
Consider the integral
\begin{equation}
I_1(L,a)=\int_a^L\textd x\int_0^\infty\textd y\int_{\R^{d-1}}\textd z\,\frac1{[(x+y)^2+|z|^2]^{s/2}}.
\end{equation}
In the limit when $a/L\to0$ (with~$L\ge1$) when~$s<d+1$ and $|\!\log a|/\log L\to0$ when~$s=d+1$,
\begin{equation}
I_1(L,a)\sim\begin{cases}
C_1L^{d+1-s},\qquad&\text{if }d<s<d+1,
\\
C_1\log L,\qquad&\text{if }s=d+1,
\end{cases}
\end{equation}
where $C_1=C_1(d,s)\in(0,\infty)$.
\end{mylemma}

\begin{proofsect}{Proof}
Scaling $z$ by $x+y$ yields
\begin{equation}
I_1(L,a)=\tilde C_1\,\int_a^L\textd x\int_0^\infty\textd y\,(x+y)^{d-1-s}
\end{equation}
where
\begin{equation}
\tilde C_1=\int_{\R^{d-1}}\frac{\textd z}{[1+|z|^2]^{s/2}}.
\end{equation}
From here the result follows by direct integration.
\end{proofsect}

Now we are ready to prove the $s<d+1$ part of Proposition~\ref{total}:

\begin{proofsect}{Proof of Proposition~\ref{total}(i)}
For~$r<1$ let~$\QQ_r$ denote the integral \eqref{Integral Q} with~$x$ restricted to a cube~$S_r$ instead of~$S_1$. Let~$\tilde T_{L,a}$ denote the quantity~$T_{L,a}$ with~$K_{i,j}$ replaced by~$\tilde K_{i,j}$. A simple scaling yields
\begin{equation}
\tilde T_{L,a}=L^{2d-s}\QQ_{1-\ffrac aL}.
\end{equation}
Hence, all we need to show is that~$\QQ_r$ remains finite as~$r\uparrow1$.
This in turn boils down to the absolute convergence of the integral defining~$\QQ$. 

To show that~$\QQ<\infty$ we note that the quantity~$(L-a)^{d-1}I_1(L-a,a)$ in Lemma~\ref{lemma-I1} may be interpreted as the integral of~$|x-y|^{-s}$ over~$x\in\Lambda_{L-a}$ and over~$y$ ranging through the half-space marked by the hyperplane passing through a given side of the cube~$\Lambda_L$. This implies
\begin{equation}
\tilde T_{L,a}\le2dL^{d-1}I_1(L-a,a)
\end{equation}
and, more importantly,
\begin{equation}
\QQ\le 2d I_1(1,0).
\end{equation}
By Lemma~\ref{lemma-I1} and the Monotone Convergence Theorem, $I_1(1,0)<\infty$ when ~$s<d+1$.
\end{proofsect}

The proof of the critical case, $s=d+1$, is more subtle. The following lemma encapsulates the calculations that are needed on top of those in Lemma~\ref{lemma-I1}:

\begin{mylemma}
\label{lemma-I2}
Let~$s\in(d,d+1]$ and consider the integral
\begin{equation}
I_2(L,a)=\int_a^L\!\textd x\int_a^L\!\textd y\int_0^\infty\!\textd\tilde x\int_0^\infty\!\textd\tilde y\int_{\R^{d-2}}\!\textd z\,\,\frac1{[(x+\tilde x)^2+(y+\tilde y)^2+|z|^2]^{s/2}}.
\end{equation}
There exists~$C_2=C_2(d,s)<\infty$ such that for $L\gg a\gg1$,
\begin{equation}
I_2(L,a)\le C_2L^{d+2-s}.
\end{equation}
\end{mylemma}

Similarly to the quantity $I_1(L,a)$ in Lemma~\ref{lemma-I1}, the integral~$I_2(L,a)$ may be interpreted as the total interaction between the square~$(-L,-a)\times(-L,-a)$ in the $(x,y)$-plane and the quarter-space in~$\R^d$ containing all points with positive~$x$ and~$y$ coordinates.

\begin{proofsect}{Proof of Lemma~\ref{lemma-I2}}
Applying the bound
\begin{equation}
(x+\tilde x)^2+(y+\tilde y)^2+|z|^2\ge x^2+\tilde x^2+y^2+\tilde y^2+|z|^2
\end{equation}
and scaling $z$ by the root of $x^2+\tilde x^2+y^2+\tilde y^2$ we get
\begin{equation}
I_2(L,a)\le O(1)
\int_a^L\!\textd x\int_a^L\!\textd y\int_0^\infty\!\textd\tilde x\int_0^\infty\!\textd\tilde y\,\, \bigl[x^2+\tilde x^2+y^2+\tilde y^2\bigr]^{\frac{d-2-s}2}
\end{equation}
Writing
\begin{equation}
r^2=x^2+y^2\quad\text{and}\quad\rho^2=\tilde x^2+\tilde y^2
\end{equation}
we pass to the polar coordinates in both pairs of variables---with~$\rho\in(0,\infty)$ and, as an upper bound, $r\in(\ffrac a2,2L)$---yielding the result
\begin{equation}
I_2(L,a)\le O(1)\int_{a/2}^{2L}\!\textd r\,r\int_0^\infty\!\textd\rho\,\,\rho[r^2+\rho^2]^{\frac{d-2-s}2}=O(1)\int_{a/2}^{2L}\textd r\,r^{d+1-s}.
\end{equation}
Here we scaled~$\rho$ by~$r$ and integrated $\rho$ out to get the last integral.
Since~$s\le d+1$, the integral over~$r$ is order~$L^{d+2-s}$.
\end{proofsect}

\begin{proofsect}{Proof of Proposition~\ref{total}(ii)}
In this case we cannot simply set~$a=0$ and apply scaling. Notwithstanding, we still have the bound
\begin{equation}
\tilde T_{L,a}\le2dL^{d-1}I_1(L-a,a).
\end{equation}
By Lemma~\ref{lemma-I1}, we have
\begin{equation}
\label{upper-tilde-T}
\tilde T_{L,a}\le2dC_1L^{d-1}\log L\bigl[1+o(1)\bigr],\qquad L\to\infty.
\end{equation}
We claim that this bound is asymptotically sharp. Indeed, \eqref{upper-tilde-T} overcounts by including (the integral over~$y$ in) the intersection of two halfspaces---marked by two neighboring sides of~$\Lambda_L$---multiple times. In light of the aforementioned interpretation of~$I_2(L,a)$, the contribution from each such intersection is bounded by~$L^{d-2}I_2(L,a)$. By Lemma~\ref{lemma-I2}, this is at most order~$L^{d-1}$. Hence we have \eqref{LlogL} with~$A=2d C_1$.
\end{proofsect}

Theorem~\ref{thm-peierls} will require us to show that, for~$s>d+1$, the total strength of the long-range interaction through the boundary of a finite set is of order boundary:

\begin{myproposition}
\label{prop-bd-domination}
Let~$s>d+1$. Then there is a constant~$C_3=C_3(d,s)<\infty$ such that if~$\Lambda\subset\Z^d$ is finite and connected, then
\begin{equation}
\label{surface_bd}
\sum_{i\in\Lambda}\sum_{j\in\Lambda^\cc}K_{i,j}\le C_3|\partial\Lambda|
\end{equation}
where~$|\partial\Lambda|$ denotes the number of bonds with one endpoint in $\Lambda$ and the other in~$\Lambda^\cc$.
\end{myproposition}

\begin{proofsect}{Proof}
Let~$V\subset\R^d$ denote the union of unit cubes centered at the sites of~$\Lambda$. Let
\begin{equation}
W=\bigl\{y\in V^\cc\colon\dist(y,V)\ge1\bigr\}.
\end{equation}
In light of \eqref{KKtilde}, it suffices to show that, for some~$C<\infty$,
\begin{equation}
\label{surface_bd_cont}
\int_{W}\textd y\int_V\textd x\,\frac{1}{|x-y|^s}\le C\Sigma(\partial V),
\end{equation}
where~$\Sigma$ denotes the surface measure on~$\partial V$. (Indeed, we have~$\Sigma(\partial V)=|\partial\Lambda|$.) To this end we note that the function~$x\mapsto (d-s)|x|^{-s}$ is the divergence of the vector field $x\mapsto x/|x|^s$. The Gauss-Green formula thus tells us that for all~$y\in(V^\cc)^\circ$,
\begin{equation}
\int_V\textd x\,\frac1{|x-y|^s}=\frac1{d-s}\,\int_{\partial\Lambda}
\frac{\tau(x)\cdot(x-y)}{|x-y|^s}\Sigma(\textd x),
\end{equation}
where~$\tau(x)$ is the unit outer normal to the surface at point~$x$ (which is well defined $\Sigma$-a.e.\ because~$\partial V$ is piecewise smooth). But $|\tau(x)\cdot(x-y)|\le|x-y|$ and so
\begin{equation}
\int_V\textd x\,\frac1{|x-y|^s}\le\frac1{s-d}\,\int_{\partial\Lambda}
\frac{1}{|x-y|^{s-1}}\Sigma(\textd x)
\end{equation}
But $s>d+1$ ensures that~$y\mapsto|x-y|^{-(s-1)}$ is integrable over $\{y\in\R^d\colon|y-x|\ge1\}$ and so integrating over~$y$, applying Fubini's theorem, extending the $y$-integral from $y\in W$ to $\{y\colon|y-x|\ge1\}$, and setting
\begin{equation}
C=(s-d)^{-1}\int_{\R^d}|z|^{1-s}1_{\{|z|\ge1\}}\,\textd z,
\end{equation}
we get~\eqref{surface_bd_cont}.
\end{proofsect}

\section{Proofs of main results}
\label{sec4}\noindent
Here we will prove the results from Sect.~\ref{sec2}; we begin with Theorem~\ref{main theorem}.
In our efforts to rule out that $m_{\star} > 0$, it is useful to have a definite state that exhibits the magnetization.  Our choice will be the limit of states at positive external field that are constructed on the torus.
  
\begin{mydefinition}
Let $h > 0$ and let $\langle-\rangle_{\T;h}$ denote an infinite volume state for the interaction described in \twoeqref{Hamiltonian}{K term} at inverse temperature $\beta$ and external field~$h$ that is constructed as a limit of finite volume states with toroidal boundary conditions.  We define $\langle-\rangle_{\T}$ to be any $h \downarrow0$ weak limit of the states $\langle-\rangle_{\mathbb T;h}$.  When the occasion arises, we will denote the measure associated with this state by $w_{\T}$.
\end{mydefinition}

\begin{mylemma}
The measure $w_{\T}$ is a Gibbs measure for the interaction described in \twoeqref{Hamiltonian}{K term} at inverse temperature $\beta$.  Moreover, $w_{\T}$ is translation invariant, it satisfies $\langle\sigma_{0}\rangle_{\T} = m_{\star}$ and if $m_L$ denotes the block magnetizations,
\begin{equation}
m_L  =  \frac{1}{|\Lambda_L|}\sum_{i \in \Lambda_L}\sigma_{i},
\end{equation}
then for any $\mu$ with $0 < \mu < m_{\star}$,
\begin{equation}
\label{Q}
\lim_{L\to\infty} w_{\T}(m_L > \mu)= 1.
\end{equation}
\end{mylemma}

\begin{proof}
These are standard results from the general theory of Gibbs states. Indeed, translation invariance follows by construction while the fact that~$w_{\T}$ is Gibbs is a result of the absolute summability of interactions; cf.~\cite[Corollary III.2.3]{Simon}. To compute the expectation $\langle\sigma_{0}\rangle_{\T}$ we recall that concavity of the free energy ensures that for any $h'<h<h''$ and any translation-invariant Gibbs state~$\langle-\rangle_{h}$ at external field~$h$,
\begin{equation}
-\frac{\partial f}{\partial h^+\!}(\beta,h')\le \langle\sigma_0\rangle_{h}\le-\frac{\partial f}{\partial h^-\!}(\beta,h'').
\end{equation}
The definition of~$m_\star$---and the construction of~$\langle-\rangle_{\T}$---then implies $\langle\sigma_{0}\rangle_{\T} = m_{\star}$. Finally, we claim that~$m_L\to m_\star$ in $w_{\T}$-probability, implying~\eqref{Q}. Indeed, if the random variable $m_L$ were not asymptotically concentrated, then
\begin{equation}
c_L \,:\,=w_{\T}(m_L>m_\star+\epsilon)
\end{equation}
would be uniformly positive (at least along a subsequence) for some~$\epsilon>0$. But then the DLR conditions and \twoeqref{surface-term}{pressure} would imply that, for any~$h>0$,
\begin{equation}
\label{ZZbound}
c_L\texte^{h|\Lambda_L|(m_\star+\epsilon)}\le\bigl\langle\,\texte^{h|\Lambda_L|m_L}\bigr\rangle_{\T}
=\biggl\langle\frac{Z_{\Lambda_L,h}}{Z_{\Lambda_L,0}}\biggr\rangle_{\T}
=\texte^{-|\Lambda_L|[f(\beta,h)-f(\beta,0)+o(1)]}.
\end{equation}
Hence we would conclude
\begin{equation}
f(\beta,h)-f(\beta,0)\le-(m_\star+\epsilon)h,
\end{equation}
in contradiction with~\eqref{spont-mag}.
\end{proof}

We now define the random analogue of the quantity $T_{L}$ denoted by $\mathbf T_L$.   In each configuration this quantity measures the antiferromagnetic interaction between the inside and outside of a box of scale $L$:
\begin{equation}
\label{Bb T}
\mathbf T_L =  
\sum_
{\begin{subarray}
{c} i \in \Lambda_L \\ j \in \Lambda^\cc_L
\end{subarray}}
K_{i,j}\sigma_i\sigma_j.
\end{equation}
The central estimate---from which Theorem \ref{main theorem} will be readily proved---is as follows:

\begin{myproposition}
\label{key one}
Consider the interaction described by \twoeqref{Hamiltonian}{K term} with $s\in(d,d+1]$ and $\beta\in(0,\infty)$ and let $m_{\star}$ denote the spontaneous magnetization corresponding to these parameters.  For each $\lambda\in(0,1)$  there is~$L_0<\infty$ such that for $L\ge L_0$, 
\begin{equation}
\label{HD}
\langle\mathbf T_{L}\rangle_{\T} \geq \lambda m_{\star}^2T_{L}.
\end{equation}
\end{myproposition}

To facilitate the proof we will state and prove a small lemma concerning the averaging behavior of the $K_{i,j}$'s:

\begin{mylemma}
\label{lemma-smoothing}
Let~$\ell$ and $a$ be such that~$a\gg\ell$ and let $V_1$ and~$V_2$ be two translates of~$\Lambda_\ell$ such that~$\dist(V_1,V_2)\ge a$. Then for any $\sigma\in\{+1,-1\}^{\Z^d}$ and any $i_0\in V_1$ and $j_0\in V_2$,
\begin{equation}
\biggl|\,\sum_
{\begin{subarray}
{c} i \in V_1 \\ j \in V_2
\end{subarray}}
K_{i,j}\sigma_i\sigma_j-K_{i_0,j_0}\Bigl(\sum_{i\in V_1}\sigma_i\Bigr)\Bigl(\sum_{j\in V_2}\sigma_j\Bigr)\biggr|
\le C\frac\ell aK_{i_0,j_0}|\Lambda_\ell|^2.
\end{equation}
Here~$C$ is a constant independent of~$a$, $\ell$,~$\sigma$, $i_0$ or $j_0$.
\end{mylemma}

\begin{proofsect}{Proof}
This is a simple consequence of the bound
\begin{equation}
\bigl|K_{i,j}-K_{i_0,j_0}\bigr|\le C\frac\ell aK_{i_0,j_0}
\end{equation}
which follows by (discrete) differentiation of the formula~\eqref{K term} and using the fact that the distance between~$V_1$ and~$V_2$ is at least~$a$, while the difference between the minimum and maximum separation of~$V_1$ and~$V_2$ is a number of order~$\ell$ and~$\ell\ll a$.
\end{proofsect}

\begin{proofsect}{Proof of Proposition~\ref{key one}}
For $a = a(L)$ tending to infinity in the fashion described in the proof of Lemma \ref{total}, it is sufficient to establish the inequality in \eqref{HD} with $T_{L}$ replaced by $T_{L,a}$ and~$\mathbf T_{L}$ replaced by its random analogue, $\mathbf T_{L,a}$, defined by the corresponding modification of \eqref{Bb T}.  We will need to introduce one more length scale, namely $\ell=\ell(L)$ which will also tend to infinity but in such a way that $\ffrac\ell a\to0$. We will assume that~$L$, $a$ and~$\ell$ are such that both $\Lambda_{L-a}$ and~$\Lambda_L^\cc$ may be tiled by disjoint copies of~$\Lambda_\ell$. (Technically this only proves the result for a subsequence but the extension is trivial.)

Let $V_{1}$ and $V_{2}$ denote translates of $\Lambda_{\ell}$ with $V_1 \subset \Lambda_{L -a}$ and $V_{2} \subset \Lambda_{L}^\cc$ and let us pick~$i_0\in V_1$ and $j_0\in V_2$. Let
\begin{equation}
q_\ell=w_\T(m_\ell>\mu).
\end{equation}
The following is now easily derived using Lemma~\ref{lemma-smoothing}: On the event that the average magnetization in both~$V_1$ and~$V_2$ exceeds~$\mu$ (which has probability at least~$2q_\ell-1$) the contribution of~$i\in V_1$ and~$j\in V_2$ to the random variable $\mathbf T_{L,a}$ is at least
\begin{equation}
\bigl[1+O(\ffrac\ell a)\bigr]K_{i_0,j_0}|\Lambda_\ell|^2\mu^2.
\end{equation}
On the other hand, on the complementary event (which has probability~$1-q_\ell$) the contribution can be as small as
\begin{equation}
-\bigl[1+O(\ffrac\ell a)\bigr]K_{i_0,j_0}|\Lambda_\ell|^2.
\end{equation}
This means that the blocks~$V_1$ and~$V_2$ contribute to $\langle\mathbf T_{L,a}\rangle_{\T}$ at least
\begin{equation}
\bigl[1+O(\ffrac\ell a)\bigr]K_{i_0,j_0}|\Lambda_\ell|^2\bigl[\mu^2(2q_\ell-1)-(1-q_\ell)\bigr].
\end{equation}
Finally, Lemma~\ref{lemma-smoothing} also gives
\begin{equation}
K_{i_0,j_0}|\Lambda_\ell|^2=\bigl[1+O(\ffrac\ell a)\bigr]\sum_{i\in V_1}\sum_{j\in V_2}K_{i,j}.
\end{equation}
Noting that the error $O(\ffrac\ell a)$ holds uniformly in the position of $V_1$ and~$V_2$, we may now sum over all (disjoint) translates of~$V_1$ and~$V_2$ in $\Lambda_{L-a}$ and~$\Lambda_L^\cc$, respectively, to get
\begin{equation}
\langle\mathbf T_{L,a}\rangle_{\T}\ge\bigl[1+O(\ffrac\ell a)\bigr]\bigl(\mu^2(2q_\ell-1)-(1-q_\ell)\bigr)T_{L,a}.
\end{equation}
Since we assumed $\ffrac\ell a\to0$ and $q_\ell\to1$ as~$L\to\infty$, the right-hand side exceeds $\lambda\mu^2T_L$ once~$L\gg1$.
\end{proofsect}

\begin{proofsect}{Proof of Theorem \ref{main theorem}}
By the inherent spin-reversal symmetry, an enhancement of the standard Peierls contour (de)erasement procedure
yields, for any $\kappa > 0$, 
\begin{equation}
\label{peierls}
w_{\T}(\mathbf T_{L} \geq \kappa T_{L}) \leq 
\texte^{-2\beta[\kappa T_{L} - 2dJL^{d-1}]}.
\end{equation}
Indeed, considering the probability conditioned on the configuration outside~$\Lambda_L$, we may split the energy into two parts: the energy inside $E_{\text{in}}(\sigma)$ and the energy $E_{\text{bdry}}(\sigma)$ across the boundary of~$\Lambda_L$. The important difference between these objects is that~$E_{\text{in}}$ is invariant under the (joint) reversal of all spins in~$\Lambda_L$, while~$E_{\text{bdry}}$ changes sign. Using the fact that the conditional measure has the Gibbs-Boltzmann form, and restricting the partition function in the denominator to configurations obeying $\mathbf T_{L} \leq-\kappa T_{L}$, we get
\begin{equation}
\label{4.15}
w_{\T}(\mathbf T_{L} \geq \kappa T_{L}|\sigma_{\Lambda^\cc})
\le\frac{\sum_{\sigma\colon \mathbf T_{L} \geq\kappa T_{L}}\texte^{-\beta[E_{\text{in}}(\sigma)+E_{\text{bdry}}(\sigma)]}}
{\sum_{\sigma\colon \mathbf T_{L} \leq-\kappa T_{L}}\texte^{-\beta[E_{\text{in}}(\sigma)+E_{\text{bdry}}(\sigma)]}}.
\end{equation}
Now let us reverse all spins in $\Lambda_L$ in the lower sum; this yields
\begin{equation}
\label{4.16}
w_{\T}(\mathbf T_{L} \geq \kappa T_{L}|\sigma_{\Lambda^\cc})
\le\frac{\sum_{\sigma\colon \mathbf T_{L} \geq\kappa T_{L}}\texte^{-\beta[E_{\text{in}}(\sigma)+E_{\text{bdry}}(\sigma)]}}
{\sum_{\sigma\colon \mathbf T_{L} \geq\kappa T_{L}}\texte^{-\beta[E_{\text{in}}(\sigma)-E_{\text{bdry}}(\sigma)]}}.
\end{equation}
But
\begin{equation}
E_{\text{bdry}}(\sigma)\ge\kappa T_L-2dJL^{d-1}
\end{equation}
for every $\sigma$ in these sums and so \eqref{peierls} holds pointwise for $w_\T(\mathbf T_{L} \geq \kappa T_{L}|\sigma_{\Lambda^\cc})$. Integrating over the boundary condition, we get \eqref{peierls}.

To finish the proof, we now note
\begin{equation}
\langle \mathbf T_{L} \rangle_{\T} \leq 
T_{L}\, w_{\T}(\mathbf T_{L} \geq \kappa T_{L}) + \kappa T_{L}\, 
w_{\T}(\mathbf T_{L} < \kappa T_{L}).
\end{equation}
We learned in Proposition~\ref{key one} that for any $\lambda < 1$ the left hand side is bounded below by 
$\lambda m_{\star}^2T_L$ for all $L$ large enough.  Thus we have, $\forall \kappa \in (0,1)$ and 
$\forall \lambda \in (0,1)$
\begin{equation}
\frac{\lambda m_{\star}^2 - \kappa}{(1-\kappa)}
\leq w_{\T}(\mathbf T_{L} \geq \kappa T_{L})
\end{equation}
once~$L\gg1$. But Proposition~\ref{total} tells us $T_L \gg L^{d-1}$ and so, in light of \eqref{peierls}, the $L\to\infty$ limit forces~$\lambda m_\star^2\le\kappa$. Taking~$\kappa\downarrow0$ yields $m_\star=0$ as claimed.
\end{proofsect}  

\begin{proofsect}{Proof of Corollary \ref{big-D}}
Let~$\mu$ be an arbitrary Gibbs state. A variant of the inequality in \eqref{ZZbound} tells us that, for any~$h>0$,
\begin{equation}
\mu(m_L>\epsilon|\sigma_{\Lambda_L^\cc})\le\texte^{-h|\Lambda_L|\epsilon}\,
\frac{Z_{\Lambda_L,h}(\sigma_{\Lambda_L^\cc})}{Z_{\Lambda_L,0}(\sigma_{\Lambda_L^\cc})}
\end{equation}
Since~$m_\star=0$, the ratio of the partition functions behaves like
\begin{equation}
\frac{Z_{\Lambda_L,h}(\sigma_{\Lambda_L^\cc})}{Z_{\Lambda_L,0}(\sigma_{\Lambda_L^\cc})}=\exp\{|\Lambda_L|\,[o(h)+o(1)]\}
\end{equation}
and so, choosing~$0<h\ll1$, the right-hand side decays exponentially in~$|\Lambda_L|$. An analogous derivation (involving~$h<0$) shows a bound on~$\mu(m_L<-\epsilon)$. The second part of the claim now follows by the Borel-Cantelli lemma.
\end{proofsect}

We will also finish the proof of the existence of magnetic order for~$s>d+1$:

\begin{proofsect}{Proof of Theorem~\ref{thm-peierls}}
The proof is a simple modification of the standard Peierls argument. Consider the box~$\Lambda_L$ and let~$\mu_L^+$ denote the Gibbs measure in~$\Lambda_L$ with plus boundary condition in~$\Lambda_L^\cc$. We claim that $\mu_L^+(\sigma_0=-1)\ll1$ once~$J$ and~$\beta$ are sufficiently large (in~$d\ge2$). Indeed, given a connected set~$\Lambda\subset\Lambda_L$ whose component is connected and which contains the origin, let~$\AA_\Lambda$ denote the event that~$\sigma_0=-1$ and that~$\partial\Lambda$ is the outer boundary of the connected component of~$-1$'s containing the origin. (In other words, $\Lambda^\cc$ is the unique infinite connected component in the complement of the connected component of~$-1$'s containing~$0$.) 

Given $\sigma\in\AA_\Lambda$, let $\sigma'$ be the result of flipping \emph{all} spins in~$\Lambda$ (including the $+1$'s). We have
\begin{equation}
H_{\Lambda_L}(\sigma)-H_{\Lambda_L}(\sigma')\ge 2J|\partial\Lambda|-2\sum_{i\in\Lambda}\sum_{j\in\Lambda^\cc}K_{i,j}.
\end{equation}
By Proposition~\ref{prop-bd-domination} the second term in the exponent is bounded by $C_3|\partial\Lambda|$. Letting~$J_0=C_3$ and applying the argument in \twoeqref{4.15}{4.16}, we thus get
\begin{equation}
\mu_L^+(\AA_\Lambda)\le\texte^{-2\beta(J-J_0)|\partial\Lambda|}.
\end{equation}
But $\mu_L^+(\sigma_0=-1)$ can be written as the sum of $\mu_L^+(\AA_\Lambda)$ over all connected~$\Lambda\subset\Lambda_L$ (with connected complement) containing the origin. The standard Peierls argument shows that this sum is dominated by the $\Lambda=\{0\}$ term once~$\texte^{+2\beta(J-J_0)}$ exceeds the connectivity constant for the so-called Peierls contours. It follows that~$\mu_L^+(\sigma_0=-1)\ll1$ for~$J>J_0$ and~$\beta$ sufficiently large, uniformly in~$L$. Taking the weak limit~$L\to\infty$ produces a magnetized infinite volume Gibbs measure $\mu^+$ and, by symmetry, a counterpart negatively-magnetized state~$\mu^-$.
\end{proofsect}

\section{Open problems}
\label{sec5}\noindent
We finish by some comments and a few open problems. First, the present paper shows the absence of magnetization at~$h=0$. A natural question is now as follows:

\begin{myproblem}
Let~$s\in(d,d+1]$. Characterize the values~$h\ne0$ at which the free energy is continuously differentiable in homogeneous external field~$h$.
\end{myproblem}

An answer to this question depends strongly on the precise structure of low-tempe\-rature states. In particular, if there is a rigid stripe order (see Problem~\ref{problem-stripes}) it is possible that, for some particular values of~$h$, there will be phase coexistence between different arrangements of stripes. Whether that has an effect on the continuity of the magnetization is not clear.

To move to our next problem, let us recall the main reason why the exponent~$s=d+1$ is critical for the disappearance of magnetic order: For $s\le d+1$, the gain to be obtained from the antiferromagnetic interaction ``through'' the boundary of a volume of scale~$L$ is order~$L^{2d-s}$ which---including the $\log L$ correction when~$s=d+1$---overpowers the short-range surface cost of order~$L^{d-1}$. However the short-range calculation only applies under the conditions where one envisions a surface tension, e.g.,  discrete spins. If we replace the Ising spins by, say, plane rotors, the cost due to local interactions for turning over a block now scales as~$L^{d-2}$. Various exponents will readjust accordingly. Thus we pose:

\begin{myproblem}
For the Ising spins replaced by $O(n)$-spins, and the spin-spin interactions given by the dot product, find the range of exponents~$s$ for which the spontaneous magnetization vanishes.
\end{myproblem}

The problem is interesting due to competing effects in the vicinity of the (purported) interfaces. It has been stipulated in~\cite{vanEnter} that, in these cases, magnetism will not occur for~$d<s<d+2$. See~\cite{vanEnterPRB} for some relevant calculations.

As for our next problem we note that, as already mentioned, absence of magnetism is far from ruling out other types of order, with striped states being a prime candidate. Thus we ask:

\begin{myproblem}
\label{problem-stripes}
Prove the existence of striped states at low temperatures for interactions of the type discussed in this note.
\end{myproblem}

Some mathematical progress~\cite{GLL} and a great deal of physical progress \cite{doniach,pokrovsky1,pokrovsky2,SK,schlovsky1,schlovsky2,vanderbilt,low,romans,chayes,tarjus,bak} in this direction has been made for the ground state problem. But, at present, the positive-temperature case is far from resolved.

Finally, we recall that much of our proof was based on thermodynamic arguments which, to begin with, require the existence of thermodynamics. Notwithstanding, analogous results should hold even for interactions that decay so slowly that the standard techniques ensuring the existence of the free energy fail. An instance of some genuine interest arises from Ref.~\cite{JKS}: Consider the model with the Hamiltonian as in \eqref{Hamiltonian} but with the long-range interaction term modified into
\begin{equation}
\sum_{i,j}K_{i,j}(\sigma_i-\rho)(\sigma_j-\rho).
\end{equation}
The quantity~$\rho$ plays the role of ``background charge'' density; the spin configurations are restricted to have average~$\rho$ (otherwise their energy diverges).

\begin{myproblem}
Suppose~$K_{i,j}\sim|i-j|^{-1}$ in~$d=2,3$ (and, in general, $K_{i,j}\sim|i-j|^{-s}$ with $s_d<s \leq d$ and $d \geq 2$). Prove that the free energy is differentiable in~$\rho$, at~$\rho=0$.
\end{myproblem}

On the basis of~\cite{SK2} one can infer that the lower bound, $s_d$, on the region of exponents in the previous open problem satisfies~$s_d\le d-1$. However, it is noted that, for $s=d-2$, there is a (complicated) counterexample to differentiability~\cite{Huse} so, presumably, $s_d\ge d-2$.

\section*{Acknowledgments}
\noindent
The authors wish to thank Aernout van Enter for many useful comments on the content and literature.
This research was partially supported by the grants NSF DMS-0505356 (M.B.), NSF DMS-0306167 (L.C.) and DOE DE-FG03-00ER45798 (S.K.).

\end{document}